\documentclass{article}

\usepackage{spconf}

\usepackage{url}
\usepackage{./bs_macros}
\usepackage{float}
\usepackage[labelformat=simple]{subcaption}

\usepackage{bbm}

\graphicspath{{./Fig/}}

\theoremstyle{definition}

\newtheorem{remark}{Remark}

\newtheorem{example}{Example}

\title{Nonstationary Portfolios: Diversification in the Spectral Domain}
\name{Bruno Scalzo $^{1}$, Alvaro Arroyo $^{1}$, Ljubi$\check{\text{s}}$a Stankovi\'c $^{2}$, Danilo P. Mandic $^{1}$}

\address{$^{1}$Department of EEE, Imperial College London, London, SW7 2BT, UK \\
	$^{2}$Faculty of Electrical Engineering, University of Montenegro, Podgorica, 81000, Montenegro \\
	Emails: \{bruno.scalzo-dees12, alvaro.arroyo17, d.mandic\}@imperial.ac.uk, ljubisa@ucg.ac.me
}

\begin{document}
\ninept

\maketitle

\begin{abstract}
Classical portfolio optimization methods typically determine an optimal capital allocation through the implicit, yet critical, assumption of statistical time-invariance. Such models are inadequate for real-world markets as they employ standard time-averaging based estimators which suffer significant information loss if the market observables are non-stationary. To this end, we reformulate the portfolio optimization problem in the spectral domain to cater for the nonstationarity inherent to asset price movements and, in this way, allow for optimal capital allocations to be time-varying. Unlike existing spectral portfolio techniques, the proposed framework employs augmented complex statistics in order to exploit the interactions between the real and imaginary parts of the complex spectral variables, which in turn allows for the modelling of both harmonics and cyclostationarity in the time domain. The advantages of the proposed framework over traditional methods are demonstrated through numerical simulations using real-world price data.
\end{abstract}


\section{Introduction}

The principle of \textit{diversification} has become the cornerstone of decision-making in finance and economics ever since the introduction of \textit{modern portfolio theory} (MPT) by Harry Markowitz in 1952 \cite{Markowitz1952}. The MPT suggests an optimal strategy for the investment, based on the first- and second-order moments of the asset price returns, which can be formulated as a quadratic optimization task commonly referred to as the \textit{mean-variance optimization} (MVO). 

Consider the vector, $\x(t) \in \domR^{N}$, which contains the returns of $N$ assets at a time $t$, the $i$-th entry of which is given by
\begin{equation}
	x_{i}(t) = \frac{p_{i}(t) - p_{i}(t-1)}{p_{i}(t-1)}
\end{equation}
where $p_{i}(t)$ denotes the value of the $i$-th asset at a time $t$. The MVO asserts that the optimal vector of asset holdings, $\w \in \domR^{N}$, is obtained through the following optimization problem
\begin{equation}
	\max_{\w} \;\{  \w^{\Trans}\m - \lambda \w^{\Trans}\R\w \} \label{eq:MVO}
\end{equation}
where $\m = \expect{\x} \in \domR^{N}$ is a vector of expected future returns, $\R = \cov{\x} \in \domR^{N \times N}$ is the covariance matrix of returns, and $\lambda$ is a Lagrange multiplier, also referred to as the \textit{risk aversion} parameter. In practice, it is usually necessary to impose additional constraints on the values of $\w$, for instance, to constrain the portfolio leverage.

The increasing availability of computational power has naturally made MVO a ubiquitous tool for financial practitioners, however, the validity of its underlying theory remains perhaps the most debated topic in the field to date. Among issues that make MVO unreliable in practice, a major caveat is the notorious challenge of estimating the moments, $\m$ and $\R$, of nonstationary asset price movements. It has been shown that standard time-averaging based estimators of $\m$ and $\R$ typically yield portfolios that are far from truly optimal, and hence exhibit poor out-of-sample performance \cite{Klein1976, Jobson1980, Merton1980, Jorion1991, Britten1999}. Moreover, this issue is further amplified by the well-established sensitivity of MVO to perturbations of the estimates, $\m$ and $\R$, whereby small changes in the inputs may generate portfolio holdings with vastly different compositions \cite{Michaud1989, Michaud1998, Best1991, Chopra1993, Kondor2007}. 

The information loss incurred by sample estimators in nonstationary environments can be demonstrated using von Neumann's \textit{mean ergodic theorem} \cite{vonNeumann1932} and Koopman's \textit{operator theory} \cite{Koopman1931}. Consider an idealised case whereby the asset price returns evolve in time according to $\x(t) = \mathcal{S}\x(t-1)$, with $\mathcal{S}:\domC^{N}\mapsto\domC^{N}$ denoting the unitary \textit{shift operator} in a Hilbert space. The mean ergodic theorem asserts that the sample mean approaches the orthogonal subspace of $\x(t)$, that is
\begin{equation}
\lim_{T \to \infty}\frac{1}{T}\sum_{t=0}^{T-1} \x(t) = \lim_{T \to \infty}\frac{1}{T}\sum_{t=0}^{T-1}\mathcal{S}^{t}(\x(0)) = \lim_{T \to \infty}\frac{1}{T}\sum_{t=0}^{T-1}\mathcal{P}\x(0)
\label{eq:mean_ergodic_theorem}
\end{equation}
with the boundedness property governed by $\|\mathcal{P}\x(t)\|_{2} \leq \|\x(t)\|_{2}$, which arises from the Cauchy-Schwarz inequality.

To overcome the limitations of MVO in the presence of non-stationarity, there has been an increasing interest in the use of spectral analysis techniques. While spectral analysis has a long history in econometrics \cite{Granger1964, Engle1974, Granger1983}, with applications ranging from business cycle analysis \cite{Baxter1990}, option valuation \cite{Madan1999}, empirical analysis \cite{Croux2001, Ramsey2002, Huang2003, Crowley2007, Rua2010, Rua2012}, through to causality analysis \cite{Breitung2006}, its application in portfolio optimization has been rather sparse. To this end, \textit{spectral portfolio theory} \cite{Lo2015, Lo2016, Lo2019} was recently introduced with the aim to enhance portfolio performance by allowing the investors to benefit from diversifying not only across assets but also across frequencies, whereby the cyclical components of the variance and covariance of asset returns are accounted for respectively by using the periodogram and cross-spectra \cite{Engle1974}.

Despite mathematical elegance and physical intuition, there remain issues that need to be addressed prior to a more widespread application of spectral analysis to portfolio optimization. For example, spectral estimation is an inherently complex-valued task, however, spectral measures such as the power spectral density (PSD) employed in \cite{Lo2016} are magnitude-only based and hence cannot account for the information within the phase spectrum. From the maximum entropy viewpoint \cite{Jaynes1957,Burg1967,vandenBos1971,Cover1984}, such models make an implicit, yet fundamental, assumption that the phase information, which is intrinsic to complex-valued spectral data, is uniform and thus not informative. Mathematically, this is equivalent to asserting that the variable is wide-sense stationary in the time domain \cite{Mandic2009}. Furthermore, spectral measures such as the PSD are absolute (or non-centred) spectral moments and hence cannot distinguish between the information attributed by the spectral mean from that by the spectral covariance, yet these designate respectively the harmonics and cyclostationarity in the time domain. 

To this end, we formulate a spectral portfolio theory using a class of spectral estimators for nonstationary signals, whereby the harmonic and cyclostationary time-domain signal properties are designated respectively by the mean and covariance of the associated spectral representation. Unlike existing methods, the proposed spectral portfolio framework is intrinsically complex-valued and thus benefits from augmented complex statistics \cite{Naseer1993, Schreier2003, Mandic2009} in order to allow for a precise description of the interaction between the real and imaginary parts of complex spectral variables, and thus of the time-phase alignment. In this way, the proposed approach is shown to enable creation of time-varying capital allocation schemes. The advantages of the proposed framework over traditional methods are demonstrated through simulations based on real-world price data.

\section{A Class of Nonstationary Signals}

We begin by consider a real-valued signal, $\x(t) \in \domR^{N}$, which admits the following time-frequency expansion \cite{Loeve1977, Schreier2003}
\begin{equation}
\label{eq:spectral_process_expansion}
\x(t) = \int_{-\infty}^{\infty} e^{\jmath \omega t} \bbx(t,\omega) \, d \omega
\end{equation}
where $\bbx(t,\omega) \in \domC^{N} $ is the realisation of a random spectral process at an angular frequency, $\omega$, and time instant, $t$. The Hermitian symmetry, $\bbx^{\ast}(t,\omega) = \bbx(t,-\omega)$, holds so that $\x(t)$ is real-valued.

To cater for a broad variety of deterministic and stochastic time-domain signals, the spectral process is assumed to be multivariate general complex Gaussian distributed \cite{vandenBos1995}, i.e. $\bbx(t,\omega)$ follows the linear model
\begin{equation}
\bbx(t,\omega) = \bbm(\omega) + \bbs(t,\omega)
\end{equation}
where $\bbs(t,\omega) \in \domC^{N}$ is a zero-mean stochastic process, while the \textit{spectral mean}, $\bbm(\omega) \in \domC^{N}$, defined as
\begin{equation}
\bbm(\omega) = \expect{\bbx(t,\omega)}  \label{eq:spectral_mean}
\end{equation}
is time-invariant. The \textit{spectral covariance} and \textit{spectral pseudo-covariance} are also time-invariant and defined respectively as
\begin{alignat}{2}
	\bbR(\omega) & = \cov{\bbx(t,\omega)} && = \expect{\bbs(t,\omega)\bbs^{\Her}(t,\omega)}  \label{eq:spectral_covariance} \\
	\bbP(\omega) & = \pcov{\bbx(t,\omega)} && = \expect{\bbs(t,\omega)\bbs^{\Trans}(t,\omega)}  \label{eq:spectral_pseudocovariance}
\end{alignat}
where the bound $\|\bbP(\omega)\|_{2} \leq \|\bbR(\omega)\|_{2}$ holds, by virtue of the Cauchy-Schwarz inequality.

As with multivariate complex variables in general, the spectral process admits a compact \textit{augmented representation} of the form
\begin{equation}
\ubbx(t,\omega) = \left[ \begin{array}{c}
\bbx(t,\omega)	 \\
\bbx^{\ast}(t,\omega)
\end{array} \right] \in \domC^{2N}
\end{equation}
which compactly parametrizes the pdf of $\bbx(t,\omega)$ as follows \cite{vandenBos1995}
\begin{equation}
p(\ubbx,t,\omega) \! = \! \frac{ \exp \! \left[ \! - \frac{1}{2} \! \left( \ubbx(t,\omega) \! - \! \ubbm(\omega) \right)^{\Her} \! \ubbR^{-1} \!(\omega) \!  \left( \ubbx(t,\omega) \! - \! \ubbm(\omega) \right) \! \right] }{\pi^{N} \det^{\frac{1}{2}} ( \ubbR(\omega) )}
\end{equation}
with
\begin{alignat}{2}
	\ubbm(\omega) & = \expect{\ubbx(t,\omega)} && = \left[ \begin{array}{c}
		\bbm(\omega)	 \\
		\bbm^{\ast}(\omega)
	\end{array} \right] \\
	\ubbR(\omega) & = \cov{\ubbx(t,\omega)} && = \left[ \begin{array}{cc}
		\bbR(\omega) & \bbP(\omega) \\
		\bbP^{\ast}(\omega) & \bbR^{\ast}(\omega)
	\end{array} \right]
\end{alignat} 
being respectively the \textit{augmented} spectral mean and covariance. Therefore, $\bbx(t,\omega)$ is said to be distributed according to
\begin{equation}
\ubbx(t,\omega) \sim \CNormal{\ubbm(\omega),\ubbR(\omega)}
\end{equation}
Furthermore, if the time-frequency representations exhibit \textit{non-orthogonal bin-to-bin increments}, then it is necessary to also consider the following \textit{dual-frequency statistics} (for $\omega \neq \nu$)
\begin{alignat}{2}
	\bbR(\omega,\nu) & = \cov{\bbx(t,\omega),\bbx(t,\nu)} \! && = \! \expect{\bbs(t,\omega)\bbs^{\Her}(t,\nu)}  \label{eq:dual_frequency_cov} \\
	\bbP(\omega,\nu) & = \pcov{\bbx(t,\omega),\bbx(t,\nu)} \! && = \! \expect{\bbs(t,\omega)\bbs^{\Trans}(t,\nu)} \label{eq:dual_frequency_pcov}
\end{alignat}
which are referred to respectively as the \textit{dual-frequency spectral covariance} and \textit{dual-frequency spectral pseudo-covariance}. These exhibit the following properties
\begin{align}
	\bbR(\omega,\nu) & = \bbR^{\ast}(\nu,\omega)\\
	\bbP(\omega,\nu) & = \bbP(\nu,\omega)\\
	\|\bbP(\omega,\nu)\|_{2} \leq \|\bbR(\omega,\nu)\|_{2} & \leq \|\bbR(\omega)\|_{2}\|\bbR(\nu)\|_{2}
\end{align}
owing to the Cauchy-Schwarz inequality \cite{Schreier2003}.

\begin{remark} \label{remark:PSD}
	Notice that the spectral moments in (\ref{eq:spectral_mean})-(\ref{eq:spectral_pseudocovariance}) are \textit{centred}, which contrasts the usual spectral statistics based on the \textit{absolute} or \textit{non-centred} moments as in \cite{Lo2016}. It is therefore possible to express the standard PSD, denoted by $\tilde{\bbR}(\omega)$, in terms of the spectral mean and covariances as follows
	\begin{equation}
		\tilde{\bbR}(\omega) = \expect{\bbx(t,\omega)\bbx^{\Her}(t,\omega)} = \bbm(\omega)\bbm^{\Her}(\omega) + \bbR(\omega) \label{eq:PSD}
	\end{equation}
	This shows that the mean and covariance information become entangled when employing the absolute (non-centred) spectral statistics. This result also highlights that the power spectrum is inadequate for detecting harmonics in low signal-to-noise ratio environments, since $\|\bbR(\omega)\| \gg \|\bbm(\omega)\|^{2}$. The PSD of the harmonics would therefore be dominated by the power associated with the noise, thereby rendering the harmonic indistinguishable from the noise.
\end{remark}


The linearity property of the Fourier transform in (\ref{eq:spectral_process_expansion}) dictates that if the spectral processes are multivariate complex Gaussian distributed, that is, $\ubbx(t,\omega) \sim \CNormal{\ubbm(\omega),\ubbR(\omega)}$, then their time-domain counterpart, $\x(t)$, is also multivariate Gaussian distributed, since a linear function of Gaussian random variables is also Gaussian distributed. The signal, $\x(t)$, is thus distributed according to
\begin{equation}
\x(t) \sim \Normal{\m(t),\R(t)} \label{eq:pdf_time-varying_Gaussian}
\end{equation}
where $\m(t) \in \domR^{N}$ and $\R(t) \in \domR^{N \times N}$ are the time-varying mean vector and covariance matrix, defined respectively as
\begin{align}
	\m(t) & = \expect{\x(t)} \label{eq:time-varying_mean} \\
	\R(t) & = \cov{\x(t)} \label{eq:time-varying_cov}
\end{align}
which are a function of the introduced spectral statistics, as is shown next.

\subsection{Mean}

From (\ref{eq:time-varying_mean}), consider a statistical expectation of the spectral expansion of $\x(t)$, as in (\ref{eq:spectral_process_expansion}), to yield
\begin{equation}
\m(t) \! = \! \expect{\x(t)} \! = \!\! \int_{-\infty}^{\infty} \!\!\! e^{\jmath \omega t} \expect{\bbx(t,\omega)} d\omega \!  = \!\! \int_{-\infty}^{\infty} \!\!\! e^{\jmath \omega t} \bbm(\omega) d\omega \label{eq:spectral_expansion_mean}
\end{equation}
Therefore, the time-varying mean of $\x(t)$ is a multivariate real-valued \textit{harmonic} signal. Notice that for $\omega = 0$, the signal reduces to a multivariate DC component.

\subsection{Covariance}

Following from the relation in (\ref{eq:time-varying_cov}), and upon introducing the centred signal, $\s(t) = \x(t)-\m(t)$, consider the covariance of the spectral expansion of $\x(t)$, as in (\ref{eq:spectral_process_expansion}), to obtain
\begin{align}
	\R(t) & = \cov{\x(t)} = \expect{\s(t)\s^{\Trans}(t)} \notag\\
	& = \int_{-\infty}^{\infty}\!\int_{-\infty}^{\infty} \!\!\! e^{\jmath (\omega-\nu) t} \bbR(\omega,\nu) +  e^{\jmath (\omega+\nu) t} \bbP(\omega,\nu) \; d\omega d\nu
\end{align}
Therefore, the time-varying covariance of $\x(t)$ consists of a sum of \textit{cyclostationary} components, each modulated at an angular frequency, $\omega$. 

\begin{example}
	With reference to Remark \ref{remark:PSD}, we next demonstrate the benefits of employing the proposed centred spectral moments over the legacy absolute second-order spectral moments (power spectrum and complementary spectrum). Consider a single realisation of a univariate general nonstationary signal in Figure \ref{fig:total_signal}. The signal consists of two harmonics at different angular frequencies embedded in general cyclostationary noise (shown in Figure \ref{fig:constituents}). Observe that the signal constituents are completely \textit{identifiable} when employing the centred spectral moments in Figure \ref{fig:M_R_P}, whereby: (i) $M(\omega)$ designates the harmonics; (ii) $R(\omega)$ designates the WSS component; and (iii) $P(\omega)$ designates the degree of cyclostationarity. In contrast, the absolute second-order moment (PSD), $\tilde{R}(\omega)$, cannot distinguish between the harmonic and stochastic components, as shown in Figure \ref{fig:PSD_CSD}.
	
	
	\begin{figure}[h!]
		\centering
		\begin{subfigure}[t]{0.225\textwidth}
			\centering
			\includegraphics[width=1\textwidth, trim={0.2cm 0.2cm 0 0.2cm}, clip]{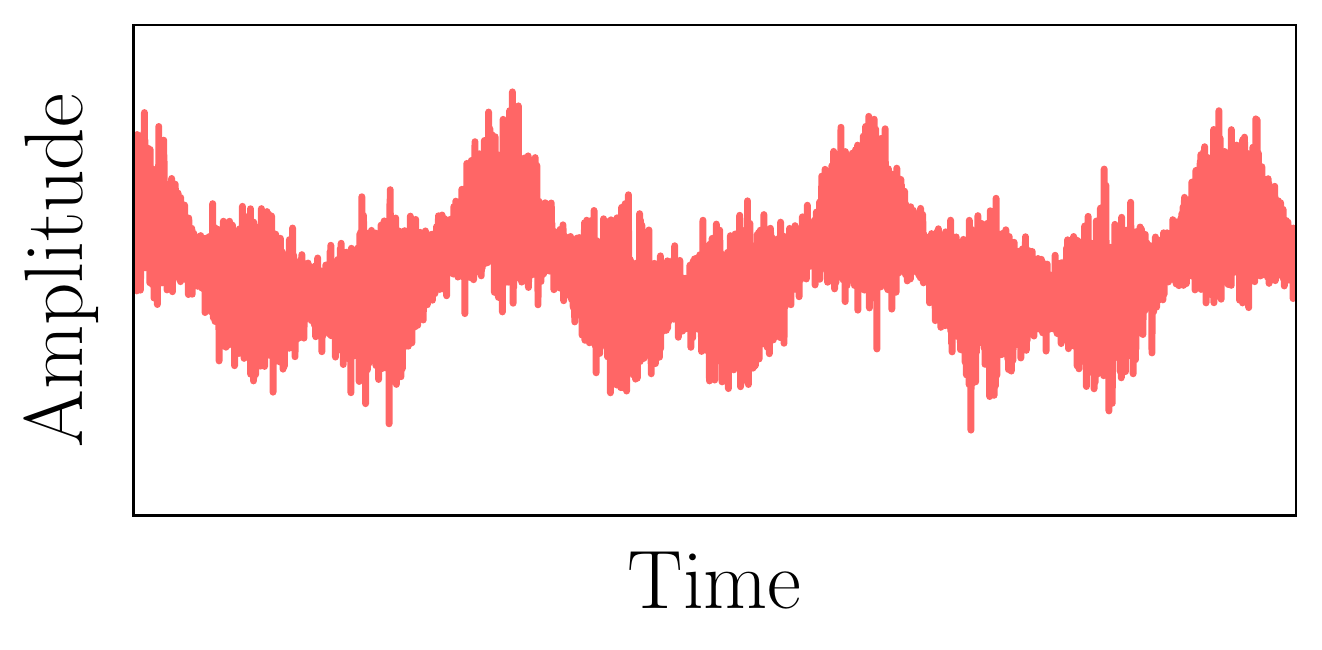} 
			\caption[]%
			{\centering{\small Nonstationary signal.}}    
			\label{fig:total_signal}
		\end{subfigure}
		\hspace{0.3cm}
		\begin{subfigure}[t]{0.225\textwidth}
			\centering
			\includegraphics[width=1\textwidth, trim={0.2cm 0.2cm 0 0.2cm}, clip]{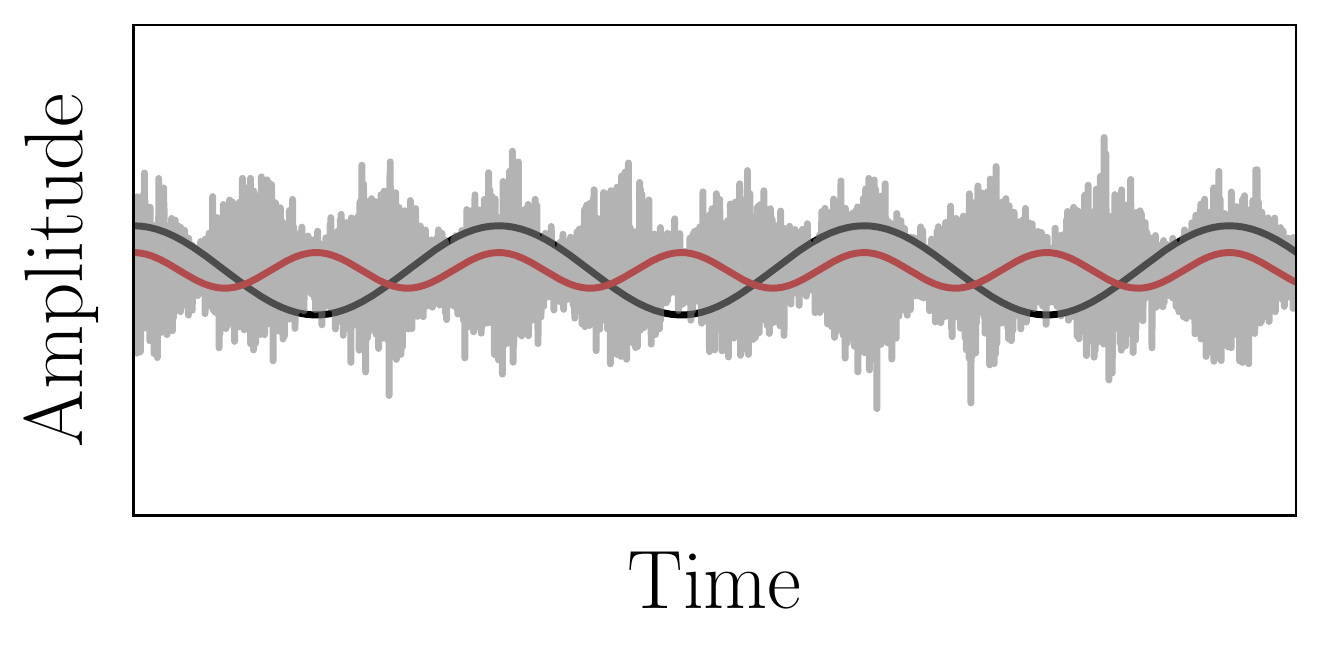} 
			\caption[]%
			{\centering{\small Constituent signals in (a).}}    
			\label{fig:constituents}
		\end{subfigure}
		\vskip\baselineskip
		\centering
		\begin{subfigure}[t]{0.225\textwidth}
			\centering
			\includegraphics[width=1\textwidth, trim={0.2cm 0.2cm 0 0.2cm}, clip]{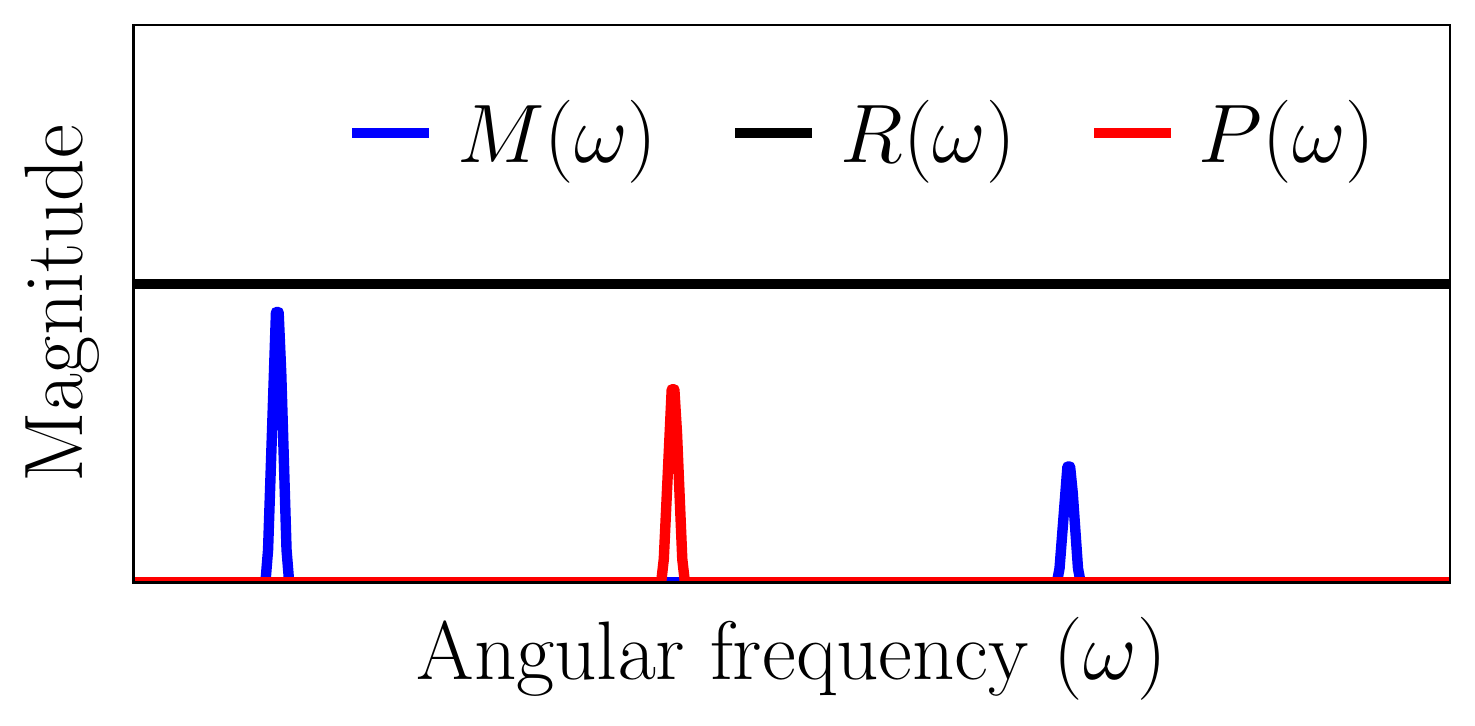} 
			\caption[]%
			{\centering{\small First- and second-order moments of the spectrum.}}    
			\label{fig:M_R_P}
		\end{subfigure}
		\hspace{0.3cm}
		\begin{subfigure}[t]{0.225\textwidth}  
			\centering 
			\includegraphics[width=1\textwidth, trim={0.2cm 0.2cm 0 0.2cm}, clip]{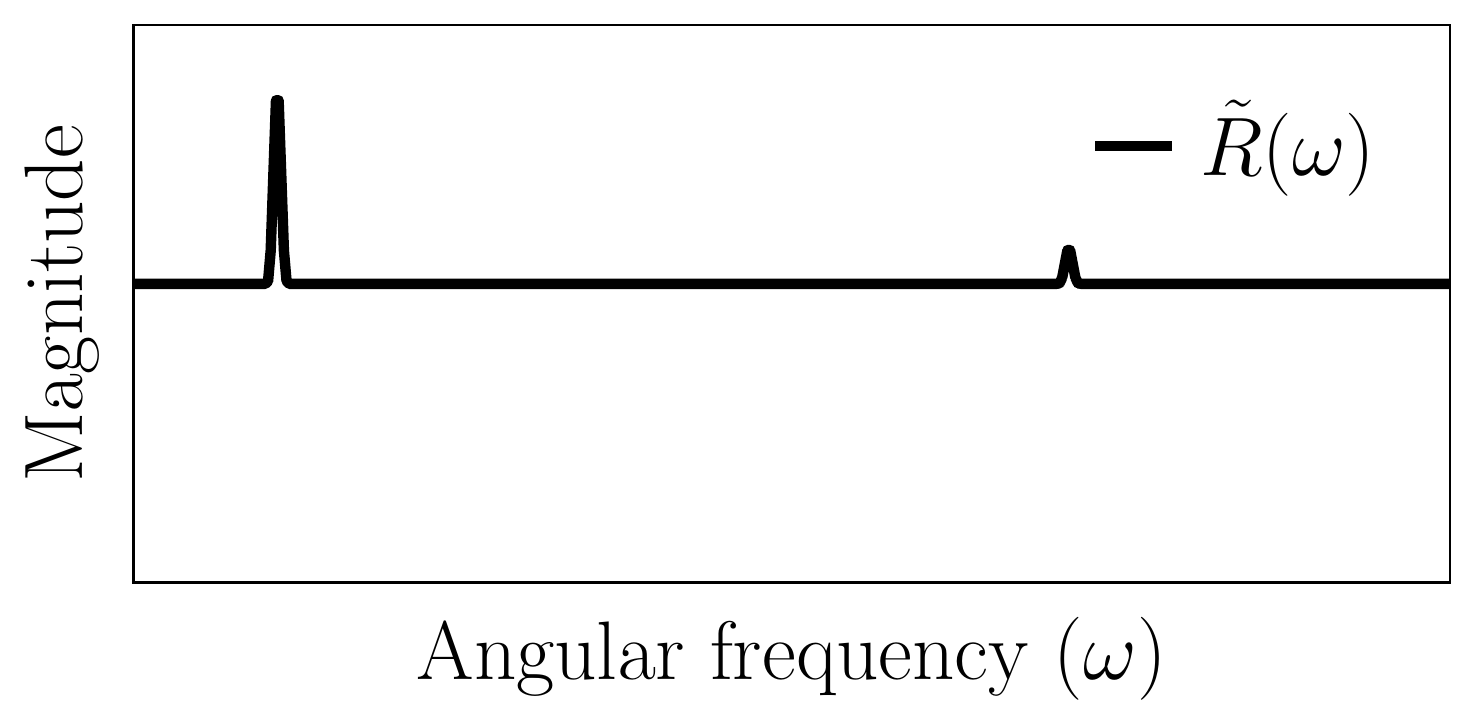} 
			\caption[]%
			{\centering{\small Absolute second-order moments of the spectrum.}}    
			\label{fig:PSD_CSD}
		\end{subfigure}
		\caption[]
		{\small Spectral analysis of a real-valued nonstationary signal. (a) A single realisation. (b) The constituents of the signal in (a). (c) The centred spectral moments. (d) The absolute spectral moments.} 
		\label{fig:nonstationary_example}
	\end{figure}
	
\end{example}



\pagebreak

\section{Compact spectral representation}


Consider a nonstationary signal which exhibits a discrete frequency spectrum, consisting of $M$ frequency bins, $\boldomega = [\omega_{1},...,\omega_{M}]^{\Trans}$. The discrete spectral expansion of $\x(t)$ in (\ref{eq:spectral_process_expansion}) therefore becomes
\begin{equation}
\x(t) = \frac{1}{\sqrt{2M}} \sum_{m=1}^{M} \left( e^{\jmath \omega_{m} t}\bbx(t,\omega_{m}) + e^{-\jmath \omega_{m} t}\bbx^{\ast}(t,\omega_{m}) \right) \label{eq:spectral process_expansion_discrete}
\end{equation}


\begin{remark}
	Unlike the conventional DFT, the normalization by the constant, $\tfrac{1}{\sqrt{2M}}$ in (\ref{eq:spectral process_expansion_discrete}), provides a rigorous mapping of coordinates from the time-domain to the time-frequency domain through a pure rotation in the complex plane, thus preserving both the desired orthogonality and norm properties \cite{Smith2007}. 
\end{remark}


To facilitate the analysis in this work, we express (\ref{eq:spectral process_expansion_discrete}) in the following compact form
\begin{equation}
\x(t)  = \uboldPhi(t,\boldomega)\ubbx(t,\boldomega) \label{eq:DFT_compact}
\end{equation}
where $\uboldPhi(t,\boldomega) \in \domC^{N \times 2MN}$ is the \textit{augmented spectral basis}, defined as
\begin{equation}
\uboldPhi(t,\boldomega)  =  \left[ \begin{array}{cc} 
\boldPhi(t,\boldomega) & \boldPhi^{\ast}(t,\boldomega) \end{array} \right]
\end{equation}
\begin{equation}
\boldPhi(t,\boldomega) = \frac{1}{\sqrt{2M}} \left[ 
\begin{array}{ccc}
e^{\jmath \omega_{1} t}\I_{N} & \cdots & e^{\jmath \omega_{M} t}\I_{N} \end{array} \right]
\end{equation}
with $\I_{N} \in \domR^{N\times N}$ being the identity matrix, and $\ubbx(t,\boldomega) \in \domC^{2MN}$ the \textit{augmented spectrum representation}, given by
\begin{equation}
\ubbx(t,\boldomega) = \left[ \begin{array}{c}
\bbx(t,\boldomega)\\
\bbx^{\ast}(t,\boldomega)
\end{array} \right], \quad 
\bbx(t,\boldomega) = \left[ 
\def\arraystretch{0.9}
\begin{array}{c}
\bbx(t,\omega_{1})\\
\vdots\\
\bbx(t,\omega_{M})
\end{array} \right] \label{eq:time-spectrum_representation}
\end{equation}
With the augmented spectrum representation, $\ubbx(t,\boldomega)$ in (\ref{eq:time-spectrum_representation}), it is now possible to jointly consider all of the dual-frequency spectral covariances in $\boldomega$ through the proposed compact formulation. To see this, consider the following probabilistic model
\begin{equation}
\ubbx(t,\boldomega) \sim \CNormal{ \ubbm(\boldomega), \ubbR(\boldomega) } \label{eq:probabilistic_spectral_model}
\end{equation}
\begin{equation}
\ubbm(\boldomega) = \expect{\ubbx(t,\boldomega)}, \quad \ubbR(\boldomega) = \cov{\ubbx(t,\boldomega)} \label{eq:augmented_expectations}
\end{equation}
where $\ubbm(\boldomega) \in \domC^{2MN}$ denotes the \textit{augmented spectral mean}, defined as
\begin{equation}
\ubbm(\boldomega) = \left[\begin{array}{c}
\bbm(\boldomega)\\
\bbm^{\ast}(\boldomega)
\end{array}\right], \quad\quad  \bbm(\boldomega) = \left[
\def\arraystretch{0.9}
\begin{array}{c}
\bbm(\omega_{1})\\
\vdots\\
\bbm(\omega_{N})
\end{array}\right]
\end{equation}
and $\ubbR(\boldomega) \in \domC^{2MN \times 2MN}$ denotes the \textit{augmented spectral covariance}, given by
\begin{align}
	\ubbR(\boldomega) & = \left[\begin{array}{cc}
		\bbR(\boldomega) & \bbP(\boldomega)\\
		\bbP^{\ast}(\boldomega) & \bbR^{\ast}(\boldomega)
	\end{array}\right] \label{eq:augmented_spectral_covariance} \\
	\bbR(\boldomega) & = \left[
	\def\arraystretch{0.9}
	\begin{array}{ccc}
		\bbR(\omega_{1}) & \cdots & \bbR(\omega_{1},\omega_{M}) \\
		\vdots & \ddots & \vdots \\
		\bbR(\omega_{M},\omega_{1}) & \cdots & \bbR(\omega_{M}) 
	\end{array}\right] \label{eq:spectrum_covariance} \\
	\bbP(\boldomega) & = \left[
	\def\arraystretch{0.9}
	\begin{array}{ccc}
		\bbP(\omega_{1}) & \cdots & \bbP(\omega_{1},\omega_{M}) \\
		\vdots & \ddots & \vdots \\
		\bbP(\omega_{M},\omega_{1}) & \cdots & \bbP(\omega_{M})
	\end{array}\right] \label{eq:spectrum_pseudo-covariance}
\end{align}

\pagebreak

To derive estimators of $\ubbm(\boldomega)$ and $\ubbR(\boldomega)$, we starting from the least squares (LS) estimate of $\ubbx(t, \boldomega)$ based on (\ref{eq:DFT_compact}), that is
\begin{equation}
\hat{\ubbx}(t, \boldomega) = \uboldPhi^{+}(t, \boldomega)\x(t) \equiv \uboldPhi^{\Her}(t, \boldomega)\x(t)
\end{equation}
with the symbol $(\cdot)^{+}$ as the pseudo-inverse operator. Next, since $\hat{\ubbx}(t, \boldomega)$ is stationary in time and hence ergodicity applies, we can simply approximate the expectation operators in (\ref{eq:augmented_expectations}) with the time-averages, to obtain the following method of moment estimators
\begin{align}
	\hat{\ubbm}(\boldomega) & = \frac{1}{T} \sum_{t=0}^{T-1} \uboldPhi^{\Her}(t, \boldomega) \x(t) \label{eq:ML_approx_mean} \\
	\hat{\ubbR}(\boldomega) & = \frac{1}{T} \sum_{t=0}^{T-1} \uboldPhi^{\Her}(t, \boldomega) \hat{\s}(t)\hat{\s}^{\Trans}(t) \uboldPhi(t, \boldomega) \label{eq:ML_approx_cov}
\end{align}
with $\hat{\s}(t) = \x(t) - \hat{\m}(t) = \x(t) - \uboldPhi(t, \boldomega)\hat{\ubbm}(\boldomega)$.


\begin{remark}
	The estimator of $\ubbm$ in (\ref{eq:ML_approx_mean}) is, in essence, the \textit{discrete Fourier transform} (DFT) of $\x(t)$. Similarly, the estimator of $\ubbR$ in (\ref{eq:ML_approx_cov}) is the power spectrum matrix of the centred variable, $\s(t)$. 
\end{remark}

\section{Spectral Portfolio Optimization}

We next derive a spectral portfolio optimization framework based on the considered class of nonstationary signals. While the standard MVO in (\ref{eq:MVO}) operates in the time-domain and with a constant capital allocation, $\w$, we instead consider optimizing the spectral content of the time-varying capital allocation, $\w(t)$. This is made possible by considering the following spectral decomposition, as in (\ref{eq:DFT_compact})
\begin{equation}
\w(t) = \boldPhi(t, \boldomega)\ubbw(\boldomega)
\end{equation}
In this way, the spectral MVO is formulated as
\begin{align}
    \max_{\ubbw(\boldomega)} \quad & \{\ubbw^{\Her}(\boldomega)\ubbm(\boldomega) - \lambda \, \ubbw^{\Her}(\boldomega)\ubbR(\boldomega)\ubbw(\boldomega)\} \label{eq:spectral_MVO} \\
    s.t. \quad & \ubbw^{\Her}(\boldomega)\ubbR(\boldomega)\ubbw(\boldomega) = \sigma_{0}^{2} \notag
\end{align}
whereby we maximise the mean portfolio return based on the spectral mean, while constrain the variance of the portfolio to a target level, $\sigma_{0}^{2}$, based on the spectral covariance. Upon inspecting the stationary points of the objective function in (\ref{eq:spectral_MVO}), we obtain the Lagrangian multiplier
\begin{equation}
    \lambda = \frac{1}{2\sigma_{0}}\sqrt{\ubbm^{\Her}(\boldomega)\ubbR^{-1}(\boldomega)\ubbm(\boldomega)}
\end{equation}
which, in turn, yields the optimal augmented spectral weights
\begin{equation}
    \ubbw_{opt}(\boldomega) = \frac{1}{2\lambda}\ubbR^{-1}(\boldomega)\ubbm(\boldomega) = \frac{\sigma_{0} \,  \ubbR^{-1}(\boldomega)\ubbm(\boldomega)}{\sqrt{\ubbm^{\Her}(\boldomega)\ubbR^{-1}(\boldomega)\ubbm(\boldomega)}}
\end{equation}
The optimal time-varying capital allocation can finally be retrieved through the augmented spectral basis, as in (\ref{eq:DFT_compact}), to yield 
\begin{equation}
    \w_{opt}(t) = \boldPhi(t, \boldomega)\ubbw_{opt}(\boldomega)
\end{equation}

\begin{remark}
In contrast to the standard MVO in (\ref{eq:MVO}), the proposed spectral portfolio framework allows for optimal time-varying capital allocation schemes. In this way, the investor is better positioned to exploit seasonal trends of asset prices, designated by the spectral mean, $\ubbm(\boldomega)$, and seasonal variations of the correlation between asset price movements, designated by the spectral covariance, $\ubbR(\boldomega)$.
\end{remark}


\section{Simulations}

\vspace{-3mm}

\begin{figure}[h!]
	\centering
	\begin{subfigure}[t]{0.4\textwidth}
		\centering
		\includegraphics[width=1\textwidth]{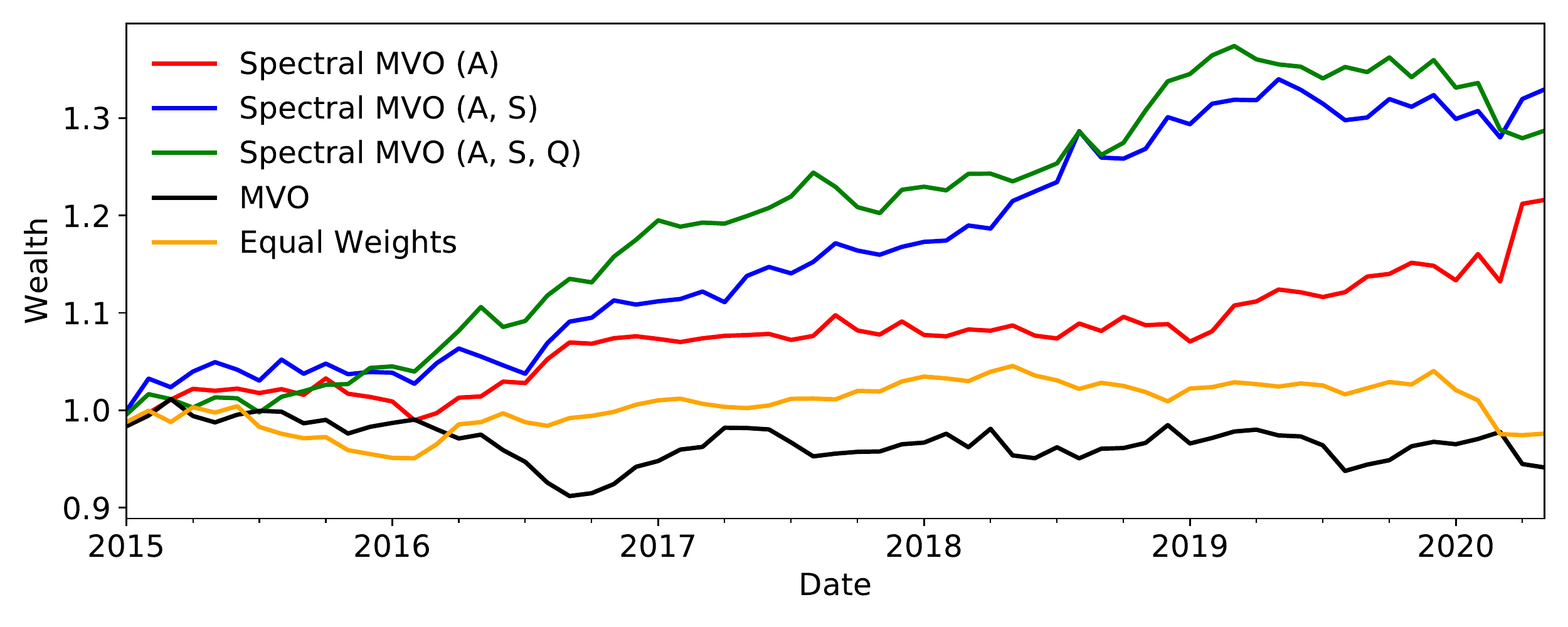}
		\caption[]{Out-of-sample performance.}
	\end{subfigure}
	\begin{subfigure}[t]{0.5\textwidth}
		\vspace{2mm}
		\centering
		\scriptsize
		\renewcommand{\arraystretch}{1}
		\setlength{\tabcolsep}{3pt}
		\begin{tabular}[H!]{| c | c | c | c | c | }
			\hline
			\makecell[c]{Spectral MVO\\ (A)} &  \makecell[c]{Spectral MVO\\ (A, S)} & \makecell[c]{Spectral MVO\\ (A, S, Q)} &  MVO &  EW \\
			\hline
			$0.91$ &  $1.31$ &  $1.21$  &  $-0.28$  &  $-0.14$  \\
			\hline
		\end{tabular}
		\caption{Annualised out-of-sample Sharpe ratios.}
	\end{subfigure}
	\begin{subfigure}[t]{0.4\textwidth}
		\centering
		\includegraphics[width=1\textwidth]{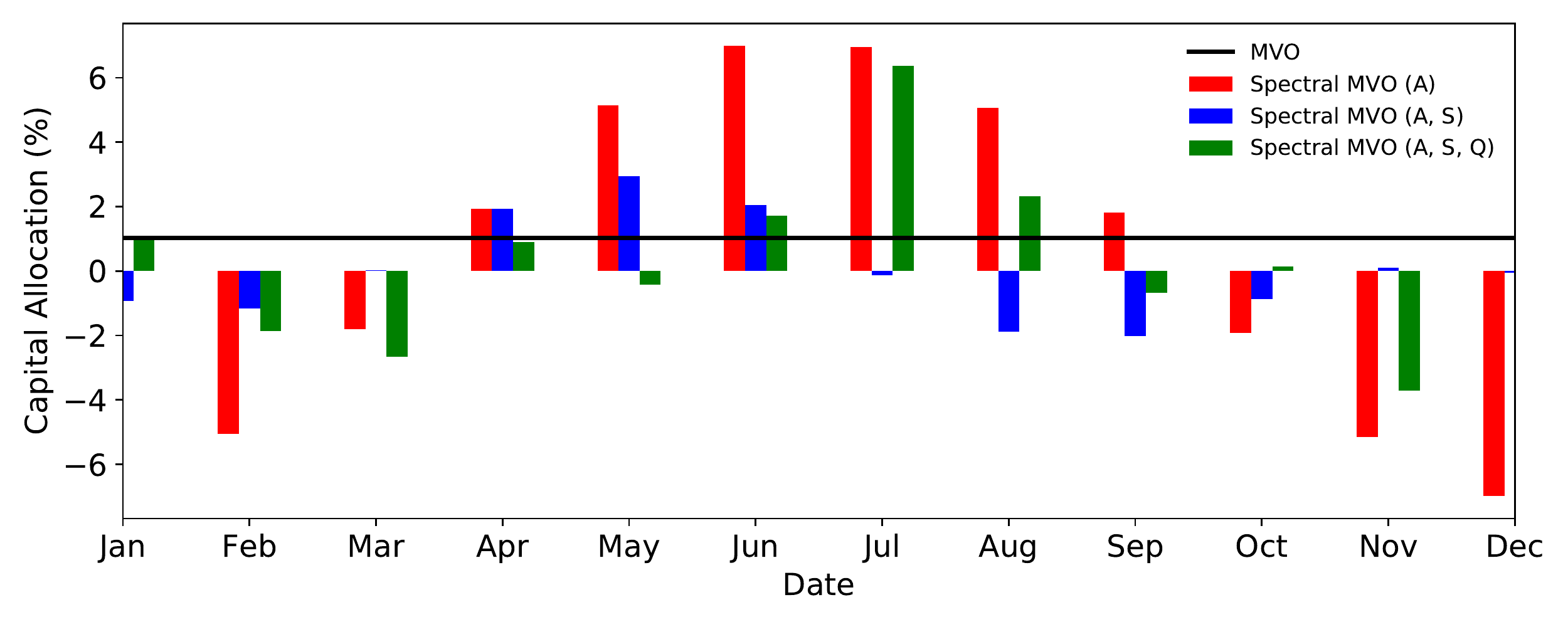}
		\caption[]{\small Allocation to gold futures by month of the year.}
	\end{subfigure}
	\vspace{-2mm}
	\caption[]{\small Investment performance for the standard MVO and the proposed spectral MVO, with varying frequency spectra, $\boldomega$. The target portfolio volatility, $\sigma_{0}$, was set to $1\%$ per annum.} 
	\label{fig:backtest}
\end{figure}

The performance of the proposed spectral MVO was investigated using monthly historical price data comprising of the $23$ commodity futures contracts constituting the Bloomberg Commodity Index, in the period 2010-01-01 to 2020-05-01. The data was split into: (i) the \textit{in-sample} dataset (2010-01-01 to 2014-12-31) which was used to estimate the spectral moments, $\ubbm(\boldomega)$ and $\ubbR(\boldomega)$, and to compute the optimal spectral weights, $\ubbw(\boldomega)$; and (ii) the \textit{out-sample} data (2015-01-01 to 2020-05-01), used to objectively quantify profitability of the asset allocation strategies. For simplicity, the frequencies chosen for $\boldomega$ corresponded to periodicities of $1$ year (A), $6$ months (S) and $3$ months (Q) ($1$ business quarter). The standard equally-weighted (EW) and MVO (MVO) portfolios were also simulated for comparison purposes, with the results displayed in Fig. \ref{fig:backtest}. 

Observe from Fig. \ref{fig:backtest} (a)--(b) that the proposed spectral MVO consistently delivered greater returns than the standard EW and MVO portfolios in the out-of-sample dataset, thereby attaining a higher \textit{Sharpe ratio}, i.e. the ratio of the mean to the standard deviation of portfolio returns. Fig. \ref{fig:backtest} (c) illustrates that by accounting for the augmented spectral information, the portfolio was better positioned to exploit time-dependent dynamics in the market, which contrasts classical approaches that assume a constant optimal allocation.

\vspace{-3mm}

\vspace{-5mm}

\section{Conclusions}

\vspace{-3mm}

A spectral portfolio theory has been introduced which employs augmented complex statistics in order to account for the full interaction between the real and imaginary parts of the complex spectra of asset price movements. This has been shown to enable the optimal capital allocation to be time-varying, which allows for the modelling of both harmonics and cyclostationarity in asset returns. Simulations have demonstrated the advantages of the proposed framework over conventional portfolio techniques, including a full utilization of the variation of the mean and covariance of asset returns in time.

\pagebreak


\bibliography{./Bibliography}
\bibliographystyle{IEEEtran}

\end{document}